\pdfoutput=1
\documentclass[aps, prb, reprint, superscriptaddress]{revtex4-2}

\usepackage{graphicx}
\usepackage{dcolumn}
\usepackage{bm}
\usepackage{siunitx}
\usepackage{amsmath}

\begin{document}

\title{Doubly resonant second-harmonic generation of a vortex beam from a bound state in the continuum}

\author{Jun Wang}
\email[]{jun.wang@epfl.ch}
\affiliation{Institute of Physics, \'Ecole Polytechnique F\'{e}d\'{e}rale de Lausanne (EPFL) (Switzerland)}
\author{Marco Clementi}
\affiliation{Dipartimento di Fisica, Universit\`a di Pavia (Italy).}
\author{Momchil Minkov}
\affiliation{Department of Electrical Engineering, Stanford University (USA)}
\author{Andrea Barone}
\affiliation{Dipartimento di Fisica, Universit\`a di Pavia (Italy).}
\author{Jean-François Carlin}
\author{Nicolas Grandjean}
\affiliation{Institute of Physics, \'Ecole Polytechnique F\'{e}d\'{e}rale de Lausanne (EPFL) (Switzerland)}
\author{Dario Gerace}
\affiliation{Dipartimento di Fisica, Universit\`a di Pavia (Italy).}
\author{Shanhui Fan}
\affiliation{Department of Electrical Engineering, Stanford University (USA)}
\author{Matteo Galli}
\affiliation{Dipartimento di Fisica, Universit\`a di Pavia (Italy).}
\author{Romuald Houdré}
\affiliation{Institute of Physics, \'Ecole Polytechnique F\'{e}d\'{e}rale de Lausanne (EPFL) (Switzerland)}

\date{\today}




\begin{abstract}

Second harmonic generation in nonlinear materials can be greatly enhanced by realizing doubly-resonant cavities with high quality factors. However, fulfilling such doubly resonant condition in photonic crystal (PhC) cavities is a long-standing challenge, because of the difficulty in engineering photonic bandgaps around both frequencies. Here, by implementing a second-harmonic bound state in the continuum (BIC) and confining it with a heterostructure design, we show the first doubly-resonant PhC slab cavity with $2.4\times10^{-2}$ W$^{-1}$ conversion efficiency under continuous wave excitation. We also report the confirmation of highly normal-direction concentrated far-field emission pattern with radial polarization at the second harmonic frequency. These results represent a solid verification of previous theoretical predictions and a cornerstone achievement,  not only for nonlinear frequency conversion but also for vortex beam generation and prospective nonclassical sources of radiation.
\end{abstract}

\maketitle

\section{Introduction}

The interaction of electromagnetic radiation with an optical medium may lead to a multitude of nonlinear processes due to the intrinsic material susceptibilities, such as the $\chi^{(2)}$ tensor. The latter is responsible, e.g., for frequency down (spontaneous parametric down conversion, SPDC) or up (second-harmonic generation, SHG) conversion, among other frequency mixing processes. Enhancing such processes is desirable for a number of applications including nonlinear spectroscopy \cite{Heinz1982}, frequency doubling of infrared laser sources to the visible or near-infrared, biosensing \cite{Campagnola2003,Estephan2010},
quantum frequency conversion  \cite{Tanzilli2005,Rakher2010,Zaske2012}, and generation of non-classical radiation  \cite{Gerace2013,Gerace2014,Caspani2017,Marino2019}. 

In SHG and SPDC, in particular, the conversion efficiency can be strongly enhanced in doubly resonant cavities, i.e., simultaneously supporting resonant modes at either first- (FH) or second-harmonic (SH) frequencies, respectively \cite{drummond1980non, paschotta1994nonlinear, berger1997, rodriguez2007chi}. In such cavities, nonlinear processes are enhanced by the quality ($Q$) factors of the two modes (i.e., increased temporal confinement), as well as by the spatial field confinement. The latter condition additionally requires that a large spatial overlap between the two fields is fulfilled,  which generalizes the phase matching condition in propagating geometries \cite{lin2016}. Doubly-resonant cavities have been proposed and experimentally demonstrated in dual period Bragg mirrors \cite{liscidini2006,Rivoire2011,buckley2014}, birefringently phase-matched waveguides \cite{scaccabarozzi2006}, and geometric dispersion-tuned  micro-ring resonators \cite{pernice2012, bruch2018}. Photonic crystal (PhC) defect cavities patterned in two-dimensional (2D) slabs, which allow for very tight field confinement in purely dielectric resonators, have been shown to produce significant SHG enhancement in a singly-resonant regime at FH \cite{McCutcheon2007, rivoire2009, Galli2010, Yamada2014, Buckley2014a, Zeng2015, mohamed2017, Song2019, Clementi2019}. However, implementing a doubly-resonant condition in PhC slab cavities is a longstanding challenge, because the SH frequency range generally lies entirely inside the light cone of the cladding materials, such that efficient confinement in the out-of-plane direction is prevented, not to mention the difficulty of engineering photonic bandgaps around both frequencies to favor the in-plane confinement.

Recently, a theoretical design strategy based on engineering a bound state in the continuum (BIC) and then employing heterostructure photonic confinement opened up a new path for doubly-resonant cavities on PhC slabs \cite{minkov2019doubly}. The BIC is a resonance of an infinitely in-plane extended PhC that lies inside the light cone but is nevertheless non-radiative, either because of symmetry protection or because of destructive interference between different radiation channels \cite{hsu2016, hsu2013observation, zhen2014}. In addition, the in-plane light confinement at SH frequency is ensured by a heterostructure design, which provides a confined mode even in the absence of a photonic bandgap \cite{ge2018}. While the in-plane confinement introduces losses into the light cone for the BIC, this SH confined mode still keeps a reasonably high Q-factor \cite{minkov2019doubly}. The confinement at FH frequency is the same as in a conventional PhC approach, i.e. exploiting total internal reflection in the out-of-plane direction and photonic bandgap confinement in the slab plane  \cite{mohamed2017}.

The BIC effect in the heterostructure cavity is highly interesting beyond simply being a means to achieve a long-lived mode at SH. In fact, BICs in PhC slabs are associated with a topological charge and are robust to structural modifications \cite{zhen2014}. Strikingly, this topological charge manifests itself in the far-field radiation in the vicinity of a BIC in momentum space. Specifically, it was shown that the far-field must be linearly polarized and that, since the emission goes to zero at the BIC, the polarization angle must have a nontrivial winding around it \cite{zhen2014}. This has also been demonstrated experimentally \cite{Doeleman2018}. In Ref.~\cite{zhen2014} it was also proposed that this effect could be used to create vortex beam lasing \cite{Zhan2009}, which can find applications in, e.g., optical trapping \cite{Ng2010}, light focusing and imaging \cite{Zhan2009, Kitamura2012}, and communications \cite{Wang2016}. 

Here we make use of such doubly-resonant PhC cavity design strategy and experimentally demonstrate highly efficient SHG in a small-footprint device fabricated in epitaxially grown highly nonlinear wide bandgap GaN material. We also confirm that the SHG signal is a radially polarized vortex beam, a consequence of the BIC, and is highly concentrated around the normal direction. This allows for an extremely high collection efficiency even with a small numerical aperture of the collecting lens, which is different from previous SHG realizations in singly-resonant PhC cavities \cite{rivoire2009,Galli2010,Buckley2014a,mohamed2017}.

\section{Experiment}\label{sec:experiment}

\subsection{Cavity design}\label{sec:cavity_design}
\begin{figure}[tb!]
	\centering
	\includegraphics[width=1\columnwidth]{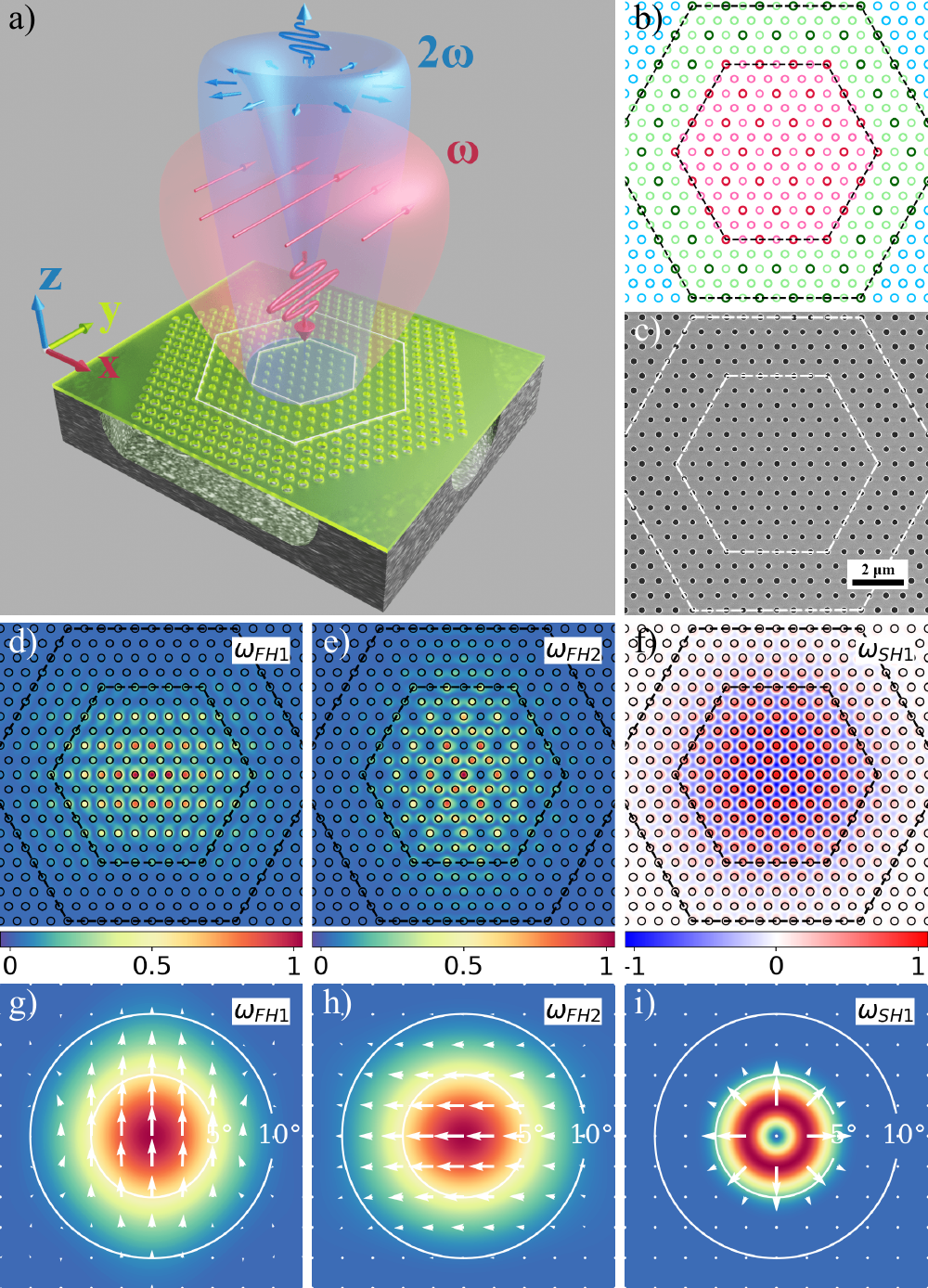}
	\caption{(a) Diagram of the SHG in a doubly resonant PhC slab cavity under linearly polarized beam excitation. (b) Design of a doubly-resonant cavity with parameters: $a=650$ nm, $N_c=6$, $N_t=4$, $N_o=14$, $r_c=130$ nm, $r_t=137$ nm, $r_o=150$ nm, $r_{c,inj}=140$ nm, $r_{t,inj}=147$ nm. The core, transition and outer regions are in red, green and blue, respectively. Injectors in core and transition regions are emphasized with crimson and dark green. (c) SEM image of the cavity. Core and transition regions are outlined by white dashed hexagons. (d-i) Near field and far field patterns of the designed cavity given by simulation (see supplement 1 for high order modes). (d-e) Intensity of the squared electric field, $|E_x^2+E_y^2|$, at the center of the slab for the first two degenerate resonant modes at FH. (f) Electric field amplitude, $\operatorname{Re}\{E_z\}$, at the center of the slab for the first resonant mode at SH. (g-h) Far-field emission pattern for the first two degenerate resonant modes at FH. The color map shows the norm of the electric field, while the overlaid arrows show the field directions. (i) Far-field emission pattern for the first resonant mode at SH.}
	\label{fig:figure_01}
\end{figure}

The basic structure of the cavity is a 2D PhC made from a hexagonal lattice of air holes in a slab, where the lattice constant is $a$, the air hole radius is $r$, the slab thickness is $d$ and the refractive index is $n$ (Fig.~\ref{fig:figure_01}~a) \cite{minkov2019doubly}. A heterostructure design is introduced to the PhC slab by increasing the hole radii ($r_c$, $r_t$, $r_o$) in three concentric hexagonal regions (core, transition and outer) whose sizes are defined by the side-lengths in units of lattice constant ($N_c$, $N_t$, $N_o$) (Fig.~\ref{fig:figure_01}~b). Theoretical simulations have been performed by three-dimensional finite-difference time domain (3D-FDTD). At FH frequency, the in-plane light confinement is ensured by the photonic band bap of the outer region (Fig.~\ref{fig:figure_01}~d-e), while at SH frequency, the confinement is given by the heterostructure design without a photonic bandgap (Fig.~\ref{fig:figure_01}~f). Far-field emission pattern at FH frequency is engineered with band-folding, which slightly increases the hole radii, with lattice period $2a$, by $\Delta r_c$ and $\Delta r_t$ in the core and transition regions, respectively. This technique folds the k-vector components at the Brillouin zone edge to the $\Gamma$-point in the reciprocal space, which concentrates the emission to the normal direction of the PhC slab and thus increases the in-coupling efficiency at FH frequency \cite{combrie2009directive,Portalupi2010}. The holes with increased radii are referred to as injectors (or extractors) and their radii are $r_{c,inj}$ and $r_{t,inj}$, respectively.

The resonant modes at FH and SH were designed at wavelengths around 1550 nm and 775 nm, respectively, for which the design parameters are: $a=650$ nm, $N_c=6$, $N_t=4$, $N_o=14$, $r_c=130$ nm, $r_t=137$ nm, $r_o=150$ nm, $r_{c,inj}=140$ nm, $r_{t,inj}=147$ nm, $d=214$ nm (Fig.~\ref{fig:figure_01}~b). To account for material dispersion, in simulations we assume refractive indices $n = 2.28$ at FH and $n = 2.31$ at SH, respectively. The FH mode is predominantly TE-polarized while the SH mode is predominantly TM-polarized (with respect to the slab plane), and the two are coupled through the $xxz$ and $yyz$ components of the GaN second-order susceptibility tensor. And since the dependencies of FH and SH resonant frequencies on the PhC parameters, such as the lattice constant $a$, the hole radius $r$ and the slab thickness $d$, are different, lithographic tuning of these parameters will help to achieve the doubly-resonant condition \cite{minkov2019doubly}. In practice, the lattice constant $a$, the hole radii $r$ and the slab thickness $d$ are scanned around the target values to verify the predicted dependencies, and also to compensate for fabrication imperfections and uncertainty of the refractive index in the experiment compared with the values used in the simulation.

\subsection{Fabrication}\label{sec:fabrication}
GaN was chosen to be the slab material because of the perfect combination of a large nonlinear susceptibility and a wide bandgap that accommodates the optical transmission for both pumping wavelengths at the telecom band and SHG wavelength at visible. The fabrication process is similar to that in previous works \cite{vico2013integrated, mohamed2017}. The GaN film (around 200 nm) was grown hetero-epitaxially on Si(111) substrate along the c-axis (z-direction) by Metal-Organic Chemical Vapor Deposition (MOCVD) at typically \SI{1000}{\celsius}. The Si(111) substrate is placed such that one of the cleavage planes is along the y-direction. An Aluminum Nitride (AlN) buffer layer (around 40 nm) was grown prior to GaN with the same condition to mitigate the lattice mismatch. 

The PhC was fabricated through a 2-step ZEP-SiO$_2$ electron beam lithography (Fig.~\ref{fig:figure_01}~c). First, a layer of 100 nm SiO$_2$ was grown on the GaN surface by Plasma Enhanced Chemical Vapor Deposition (PECVD). Then a layer of around 50 nm positive resist (ZEP520A) was spin-coated on top of SiO$_2$ and was patterned by electron beam (Vistec EBPG5000+). After development, the pattern in the resist was transferred to SiO$_2$ layer to form a hard mask by Inductively Coupled Plasma Reactive Ion Etching (ICP-RIE). After removal of ZEP residue (by remover-1165), another step of ICP-RIE dedicated to III-nitride was applied to transfer the pattern from SiO$_2$ hard mask to GaN/AlN layer. Finally, after removal of residual SiO$_2$ by Hydrogen Fluoride (HF), isotropic Xenon Difluoride (XeF$_2$) gas etching was applied to under-etch the Si substrate through holes in the GaN/AlN layer and an air gap around \SI{2}{\micro\meter} was created.

\subsection{Characterization}\label{sec:characterization}

\subsubsection{Resonances and detuning}\label{sec:resonances_and_detuning}
\begin{figure}[htb]
	\centering
	\includegraphics[width=1\columnwidth]{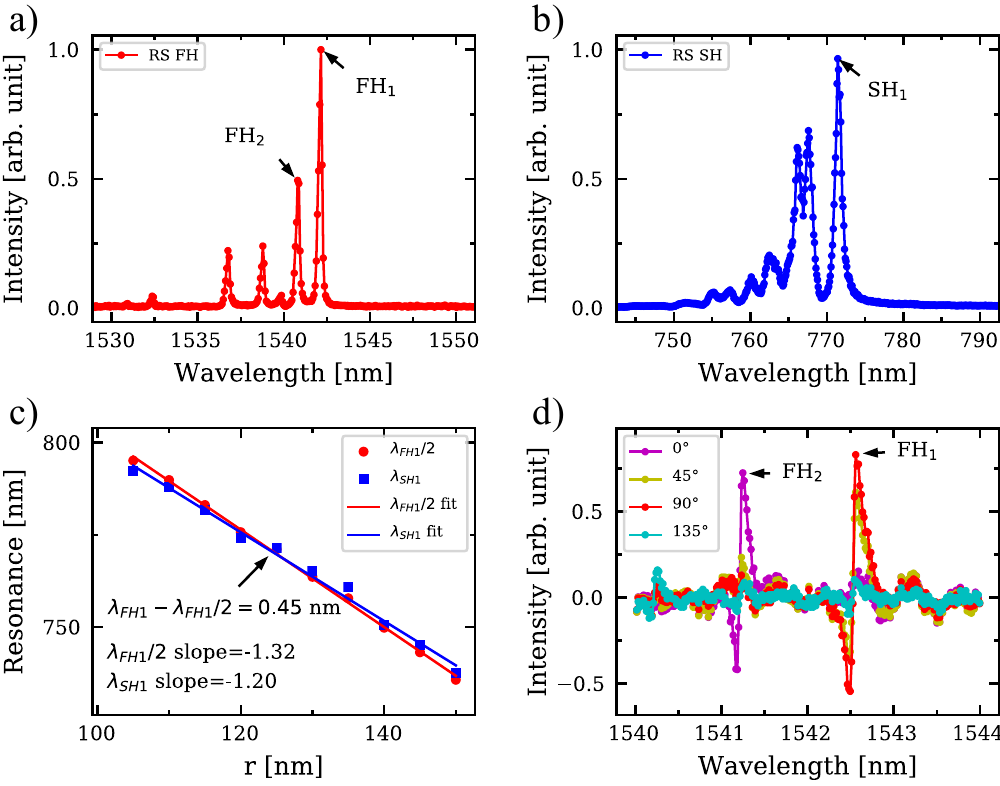}
	\caption{(a-b) Typical resonant scattering (RS) spectra at FH and SH, obtained by super-continuum broad-band excitation and spectrometer detection. (c) Dependencies of FH and SH resonances on the hole radius. $r$ is the hole radius in the core region, while the radius increments for transition and outer regions are constant. (d) FH resonant scattering spectra at different incident polarization with Fano-shaped peaks, after removal of analyzer. The first resonance peak (FH$_1$) is excited at around 90\si{\degree} (from x-axis) while the second resonance peak (FH$_2$) is excited at around 0\si{\degree}.}
	\label{fig:figure_02}
\end{figure}

Resonant scattering (RS) technique \cite{galli2009light} was employed to characterize the cavity resonances at both FH and SH wavelengths. At FH range (around 1550 nm wavelength), the cavity was excited by a linearly polarized beam at normal incidence with E-field at 45\si{\degree} from the x-axis of the cavity, knowing from simulations \cite{minkov2019doubly} that the resonant modes are linearly polarized along x or y axis (Fig.~\ref{fig:figure_01}~g-h). Typically 5 peaks are visible within 5 nm range (Fig.~\ref{fig:figure_02}~a). The peaks from long wavelength to short are indexed as FH$_1$, FH$_2$ and so on. Q-factor of $2.0 \times 10^4$ can be measured for FH$_1$, which is comparable with those measured in GaN-based L3 and H0 singly resonant cavities \cite{mohamed2017}. The theoretical quality factor of the cavity without the extractors is $1.1 \times 10^5$, while the value with $\Delta r_c = 10$ nm is $2.6 \times 10^4$. This suggests that the measured quality factor is limited by the extractor size. Smaller extractors could thus lead to higher $Q$ and conversion efficiency, since even the nominal cavity without extractors was found to have sizable far field components in the vertical direction \cite{minkov2019doubly}.  


At SH range (around 775 nm wavelength), the mode was excited with the same configuration as for FH. The quasi-TM mode at SH could be excited because of non-zero overlap with the incident E-field (Fig.~\ref{fig:figure_01}~i). A sharp and intense peak can be observed, together with some higher order peaks at smaller wavelengths (Fig.~\ref{fig:figure_02}~b). Q-factor of around 800 can be measured while the theoretical value is around 2000. Three-dimensional FDTD simulations suggest that the SH mode Q-factor is not sensitive to the extractor size. 

An important characteristic of doubly-resonant cavities is how the two resonances match simultaneously the FH and SH wavelengths. The detuning of doubly resonance can be defined as:
\begin{equation}\label{eq:definition_of_detuning}
\Delta \lambda = \lambda_\text{FH}/2 - \lambda_\text{SH},
\end{equation}
where $\lambda_\text{FH}$ and $\lambda_\text{SH}$ are the resonant wavelengths at FH and SH, respectively. The SHG conversion efficiency follows \cite{liscidini2006}:
\begin{equation}\label{eq:conversion_efficiency}
\eta_{\text{conv}}(\lambda) \propto Q_{\text{FH}}^2 Q_{\text{SH}} \mathcal{L}_{\text{FH}}^2(\lambda) \mathcal{L}_{\text{SH}}(\lambda/2),
\end{equation}
where $\lambda$ is the excitation wavelength, $Q_{\text{FH}}$ is the Q-factor of FH resonance, $Q_{\text{SH}}$ is the Q-factor of SH resonance, $\mathcal{L}_{\text{FH}}(\lambda)$ is the Lorentzian function of FH resonance, $\mathcal{L}_{\text{SH}}(\lambda)$ is the Lorentzian function of SH resonance. When exciting the cavity at FH resonance, i.e., $\lambda=\lambda_{\text{FH}}$, the detuning $\Delta \lambda$ determines the conversion efficiency $\eta_{\text{conv}}$ via $\mathcal{L}_{\text{SH}}(\lambda_{\text{FH}}/2)$, and the smaller the detuning, the higher the conversion efficiency. 

As mentioned before, the PhC parameters, such as lattice constant $a$, hole radius $r$ and slab thickness $d$, are lithographically scanned to match the doubly-resonant condition. The dependencies of $\lambda_{\text{FH}}$ on PhC hole radius $r$ is observed to be more sensitive than that of $\lambda_{\text{SH}}$, and a crossing at around the target hole radius shows up at around the target wavelength, which is in agreement with theoretical predictions \cite{minkov2019doubly} (Fig.~\ref{fig:figure_02}~c). Similar crossings on lattice constant and slab thickness are also achieved (see supplement 1).


\subsubsection{FH Cavity mode polarization}\label{sec:cavity_mode_polarization}
The polarization of the first two modes was investigated by exciting the cavity with incident beams at different polarization (Fig.~\ref{fig:figure_02}~d). The experimental setup was modified from the resonant scattering one \cite{galli2009light} such that the polarization of the incident beam could be rotated with a $\lambda/2$ waveplate and the analyzer was removed. The results show that the peak FH$_1$ is excited at around 90\si{\degree} (from x-axis of the cavity) polarization while the peak FH$_2$ at around 0\si{\degree}, in agreement with the theoretical predictions (Fig.~\ref{fig:figure_01}~g-h).

\subsubsection{Second harmonic generation}
\label{sec:second_harmonic_generation}
The second harmonic generation was investigated by exciting the cavity with a linearly polarized laser beam, at normal incidence and with the electric field at 45\si{\degree} with respect to the x-axis of the cavity. The collimated beam from a tunable continuous-wave laser source was focused on the cavity by a microscope objective (20$\times$, NA=0.4) and the coupling to cavity was optimized by fine translation of the sample in x, y and z directions using a piezoelectric stage. The SHG signal was collected through the same objective, redirected by a dichroic mirror, and detected with a Si photodetector in free space.

\begin{figure}[tb!]
	\centering
	\includegraphics[width=1\columnwidth]{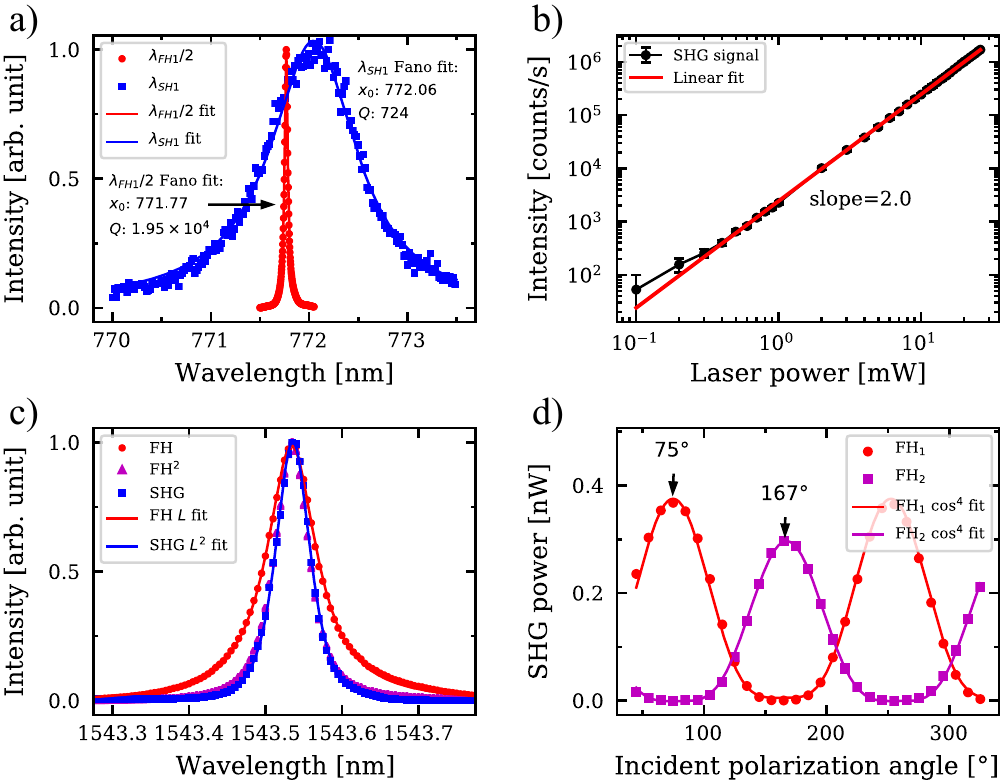}
	\caption{(a) Normalized resonant scattering intensity at FH and SH as a function of excitation wavelength, for a cavity with 0.3 nm detuning. (b) SHG intensity as a function of excitation power at resonance. (c) Normalized resonant scattering intensity at FH and SHG intensity as a function of excitation wavelength. (d) SHG signal at FH$_1$ and FH$_2$ as a function of incident polarization. (e) SH far-field emission pattern given by simulation. (f) Fourier imaging of SH far-field emission pattern. (g-j) Fourier imaging of SH far-field emission pattern with polarizer at 0\si{\degree}, 45\si{\degree}, 90\si{\degree} and 135\si{\degree},  respectively.}
	\label{fig:figure_03}
\end{figure}

\begin{table}[tb!]
\centering
\caption{Comparison of SHG in two cavities, where $\Delta\lambda$ is the detuning, $Q_{\text{FH}}$ and $Q_{\text{SH}}$ are Q-factors for FH and SH resonances, $\mathcal{L}_{\text{FH}}(\lambda_{\text{FH}}/2)$ is the amplitude of the normalized SH Lorentzian function when excitation is at FH resonance, $\eta_{\text{c}}$ is the coupling efficiency,  $P_{\text{c}}$ is the coupled power, $P_{\text{SHG}}$ is the SHG power, $\eta_{\text{conv}}$ is the conversion efficiency.}
\begin{tabular}{ccccccccc}
	\hline
	 &$\Delta\lambda$ &$Q_{\text{FH}}$ &$Q_{\text{SH}}$ &$\mathcal{L}_{\text{FH}}$ &$\eta_{\text{c}}$  &$P_{\text{c}}$ &$P_{\text{SHG}}$ &$\eta_{\text{conv}}$ \\
	 
	 &(nm) &($\times10^4$) & &($\lambda_{\text{FH}}/2$) &(\%) &(mW) &(nW) &(W$^{-1}$) \\
	 
	\hline
	1& 0.9 &1.49 &804 &0.2 &19.6 &0.541 &1.120 &3.8$\times$10$^{-3}$\\
	\hline
	2& 0.3 &1.95 &724 &0.8 &6.9  &0.125 &0.372 &2.4$\times$10$^{-2}$\\
	\hline
\end{tabular}
\label{table:table_01}
\end{table}

Two cavities are shown as examples (Table~\ref{table:table_01}): one with large detuning (0.9 nm) and the other with small detuning (0.3 nm) (Fig.~\ref{fig:figure_03}~a) (see supplement 1 for the calibration of conversion efficiency). The ratio of the two conversion efficiencies is very close to the value predicted by Equation \ref{eq:conversion_efficiency}. Moreover, the record conversion efficiency in the cavity with small detuning, $2.4\times10^{-2}$ W$^{-1}$, is 10 times larger than that of singly resonant L3 and H0 cavities \cite{mohamed2017} ($\eta_\text{conv} = 2.4 \times 10^{-3}$ W$^{-1}$, $Q=3.3 \times 10^4$), even with smaller Q-factor at FH, which confirms the great potential of this doubly-resonant PhC cavity scheme for efficient nonlinear frequency conversion.


The SHG process is ascertained by power dependent measurement: by fixing the excitation wavelength at FH$_1$, the SHG intensity scaled quadratically with the excitation power (Fig.~\ref{fig:figure_03}~b). Alternatively, by fixing the excitation power and scanning the pumping wavelength, the SHG intensity exhibited Lorentzian-squared dependence and matched perfectly with the square of FH resonant scattering intensity, which also confirmed the quadratic nature of the SHG process (Fig.~\ref{fig:figure_03}~c).

Fine polarization scan shows that the SHG intensity is proportional to $\cos^4(\theta_p)$, where $\theta_p$ is the incident polarization angle, and the curve for FH$_2$ is dephased from FH$_1$ by about 90\si{\degree} (Fig.~\ref{fig:figure_03}~d). Together with the previous resonant scattering experiment, this confirms that the polarization of mode FH$_1$ is around 90\si{\degree} (along with y axis of the cavity) while the polarization of mode FH$_2$ is around 0\si{\degree} (along with x axis of the cavity). The 180\si{\degree} period of both modes and  $\cos^4$ dependence are consistent with the assumption of linear polarization of cavity modes at FH.

\begin{figure}[ht!]
	\centering
	\includegraphics[width=1\columnwidth]{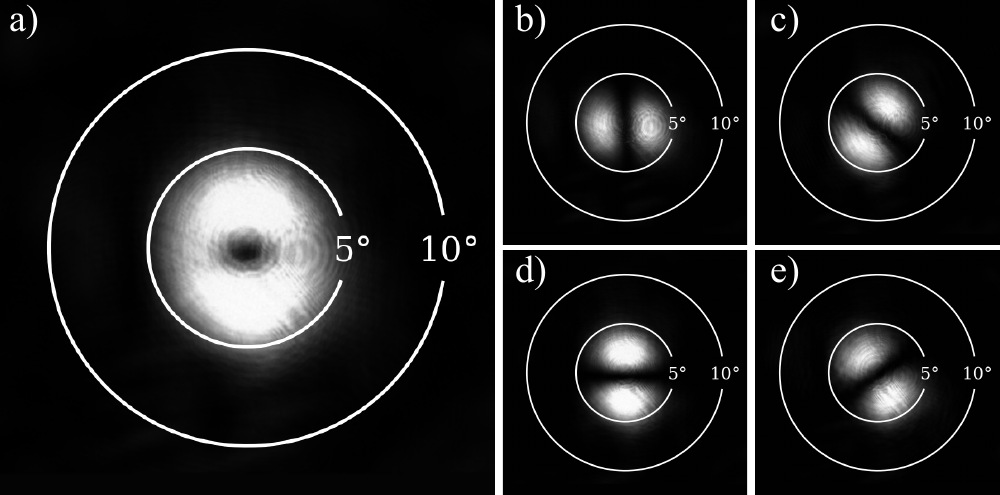}
	\caption{(a) Fourier imaging of SH far-field emission pattern. (b-e) Fourier imaging of SH far-field emission pattern with polarizer at 0\si{\degree}, 45\si{\degree}, 90\si{\degree} and 135\si{\degree},  respectively.}
	\label{fig:figure_04}
\end{figure}

\subsubsection{SH Far-field emission pattern} \label{sec:sh_far_field_emission_pattern}
Theoretical analysis and 3D-FDTD simulation predict that the far-field emission at SH frequency is a linearly polarized vector beam with a donut-shaped intensity pattern and radial polarization (Fig.~\ref{fig:figure_01}~i). This is consistent with the topological nature of BICs in momentum space \cite{zhen2014}. To verify this, the SHG far-field emission pattern was investigated by Fourier imaging. After passing through the microscope objective and being redirected by a dichroic mirror, the SHG beam was focused by a lens to form an image of the back focal plane of the objective on a CCD array, and a donut-shaped pattern was directly captured (Fig.~\ref{fig:figure_04}~a). The donut-shaped far-field pattern is concentrated within a $\pm$5\si{\degree} angle,  which is in perfect agreement with the simulation.

Moreover, when placing a polarizer film in front of the CCD array, a pattern of two lobes always remains, and its axis, which crosses the centers of the lobes, is always aligned with the polarizer, which confirms the radially polarized nature of the beam (Fig.~\ref{fig:figure_04}~b-e).

\section{Discussion}\label{sec:discussion}

The 3D-FDTD simulation predicts two degenerate modes at FH wavelength with linear polarization of the far-field emission along the x- and y-axis, respectively \cite{minkov2019doubly}, while the resonant scattering experiment and the fine polarization rotation in SHG experiment show two separated peaks with orthogonal polarization. This suggests that the two orthogonal modes are split in wavelength because of symmetry breaking, commonly observed in optical microcavities affected by disorder \cite{Hennessy2006}. Further 3D-FDTD simulations show that material birefringence and geometrical symmetry breaking (e.g., elliptical holes) can split the degeneracy of the FH mode. 


The BIC resonant mode at SH produces a linearly polarized vortex beam with radially winding polarization, which confirms the topological charge description \cite{zhen2014}. The donut-shaped far-field emission pattern is highly concentrated around the normal direction, which makes the out-coupling very efficient. Simulations suggest that the symmetry of the far-field emission pattern could be broken by material birefringence and geometry deformation. In practice, the small anisotropy with a higher intensity of the far-field emission pattern in y direction suggests systematic symmetry breaking both at material and geometrical levels, which is consistent with the split peaks at FH. 

Although the optimization of material quality and Q-factors were not the main targets during the fabrication process, the Q-factors are comparable with the predicted values, and the SHG conversion efficiencies are significantly higher than that in singly-resonant cavities made of comparable materials and through similar fabrication processes. This shows the robustness of the present design and suggests great potential for improvement in the Q-factors and conversion efficiency. Alternative doubly-resonant cavity designs have been proposed, based on topology optimization of micropost and microrings \cite{lin2016, lin2017topology}. However, these designs contain very fine structural features that make them less robust to fabrication imperfections in practice. Further design based on cylindrical dielectric structures has also been demonstrated \cite{koshelev2020}, characterized by a high degree of compactness and sustain quite high excitation powers. However, the measured Q-factors and conversion efficiencies are much lower than the ones shown in the present work, and the BIC mode is in an FH range that requires a vortex beam for efficient pumping, which is less convenient than pumping with linearly polarized beams as in the present case.

\section{Conclusion}\label{sec:conclusion}
We have experimentally demonstrated the first doubly-resonant PhC slab cavity, whose Q-factors at FH (around 1550 nm) and SH (around 775 nm) are around $2.0 \times 10^4$ and 800, respectively. An experimental SHG conversion efficiency of $2.4\times10^{-2}$ W$^{-1}$ is achieved, which is 10 times larger than the previously shown result for a singly-resonant cavity with similar material (i.e., GaN by MOCVD on Si) and processing technique. We also confirmed that the SHG emission pattern is tightly concentrated ($<\pm$5\si{\degree}) around the normal direction with donut shape and radial polarization, which originate from the BIC mode at SH.   

In this work, the implementation of BIC and heterostructure cavity put forward a practical realization to the long-standing challenge of designing PhC slab cavities fulfilling the doubly-resonant conditions. We notice that significant room for improvement is left for further developments, both at the level of nonlinear overlap factor design and experimental Q-factors, and thus SHG conversion efficiency could ultimately be increased by orders of magnitude in the future. Specific features such as tight light confinement, controllable wavelength detuning, and highly normal direction concentration of both FH and SH beams make this design an ideal platform for on-chip nonlinear light manipulation. In addition, the BIC confinement mechanism provides a natural way to generate a radially polarized vortex beam through the SHG process. 

\section{Acknowledgments}
JW and RH would like to acknowledge financial support from the Swiss National Science Foundation, through projects number 2000020-169560 and 200020-188649. DG, AB, MC, MG ackowledge the Horizon 2020 Framework Programme (H2020) through the QuantERA ERA-NET Cofund in Quantum Technologies, project CUSPIDOR, confunded from Ministero dell’Istruzione, dell’Università e della Ricerca (MIUR), and MIUR through the ``Dipartimenti di Eccellenza Program (2018-2022)'', Department of Physics, University of Pavia. MM and SF acknowledge the support of a U. S. Air Force Office of Scientific Research MURI project (Grant No. FA9550-17-1-0002).


\bibliography{bibliography}

\end{document}